\documentclass[reprint,superscriptaddress,
groupedaddress,
amsmath,amssymb,
aps,
pra,
]{revtex4-2}

\usepackage[pdftex]{graphicx}
\usepackage{dcolumn}
\usepackage{bm}
\usepackage[pdftex]{hyperref}
\usepackage{color}
\usepackage{here}

\begin{document}

\title{Quantum reservoir probing: an inverse paradigm of quantum reservoir computing for exploring quantum many-body physics}
\author{Kaito Kobayashi}
\author{Yukitoshi Motome}
\affiliation{Department of Applied Physics, the University of Tokyo, Tokyo 113-8656, Japan}
\date{\today}
\begin{abstract}
  Quantum reservoir computing (QRC) is a brain-inspired computational paradigm, exploiting natural dynamics of a quantum system for information processing.  
  To date, a multitude of quantum systems have been utilized in the QRC, with diverse computational capabilities demonstrated accordingly. 
  This study proposes a reciprocal research direction: probing quantum systems themselves through their information processing performance in the QRC framework. 
  Building upon this concept, here we develop quantum reservoir probing (QRP), an inverse extension of the QRC. 
  The QRP establishes an operator-level linkage between physical properties and performance in computing. 
  A systematic scan of this correspondence reveals intrinsic quantum dynamics of the reservoir system from computational and informational perspectives. 
  Unifying quantum information and quantum matter, the QRP holds great promise as a potent tool for exploring various aspects of quantum many-body physics. 
  In this study, we specifically apply it to analyze information propagation in a one-dimensional quantum Ising chain. 
  We demonstrate that the QRP not only distinguishes between ballistic and diffusive information propagation, reflecting the system's dynamical characteristics, but also identifies system-specific information propagation channels, a distinct advantage over conventional methods. 
\end{abstract}

\maketitle

\section{Introduction}
\label{sec:intro}

The contemporary era has witnessed an extraordinary escalation in the capabilities of artificial intelligence. 
Emulating the intricate workings of the human brain, it has revolutionized diverse domains, including image recognition and machine translation~\cite{Krizhevsky:NIPS:2012,LeCun:Nature:2015,Vaswani:NIPS:2017}. 
Nevertheless, no matter how smart it is, artificial intelligence remains constrained by the fundamental physical limitations inherent in the silicon-based substrates on which it is realized. 
Considering the substantial energy consumption and the approaching downscaling limits, the need for alternative computational paradigms has been widely recognized. 
Consequently, unconventional computing now stands as an interdisciplinary frontier in scientific exploration~\cite{Markovic:NatRevPhys:2020,Kaspar:Nature:2021,Schuman:NatComputSci:2022}. 
A leading methodology in this domain is physical reservoir computing~\cite{Maass:NeuralComp:2002,Science:Jaeger:2004,Lukosevicius:Compter:2009,Tanaka:NeuralNetw:2019,Roy:Nature:2019,Kobayashi:SciRep:2023}. 
In this brain-inspired algorithm, an input-driven dynamical system, termed a physical reservoir, performs nonlinear transformations on sequential input data. 
When the dynamics exhibit a high-dimensional internal space with pronounced nonlinearity, a simple linear transformation of read-out outcomes from the physical reservoir is sufficient to precisely generate a target output function. 
Quantum systems inherently satisfy these criteria as effective physical reservoirs, possessing intrinsic nonlinearity and an exponentially large Hilbert space. 
This has led to the development of the quantum reservoir computing (QRC) framework, which leverages quantum systems as physical reservoirs~\cite{Fujii:PRA:2017,Nakajima:PRApplied:2019}. 
Seminal proposals with spin-based implementations have demonstrated the exceptional performance of QRC~\cite{Fujii:PRA:2017,Nakajima:PRApplied:2019,Kutvonen:SciRep:2020,Pena:PRL:2021,Pena:CognComput:2023,Sannia:Quantum:2024}, later expanded to a variety of quantum systems including fermionic and bosonic networks~\cite{Ghosh:npj:2019,Ghosh:PRL:2019,Guillem:AdvQuantumTechnol:2023}, harmonic and nonlinear oscillators~\cite{Nokkala:CommPhys:2021,Angelatos:PRX:2021,Govia:PRR:2021,Dudas:npj:2023}, and Rydberg atoms~\cite{Bravo:PRXQ:2022}. 
Furthermore, recent advancements in quantum technologies have facilitated proof-of-principle experiments for the QRC across several quantum reservoir settings, such as nuclear magnetic resonance systems~\cite{Negoro:arXiv:2018} and superconducting qubits~\cite{Chen:PRA:2020,Suzuki:SciRep:2022}. 
Importantly, the diverse computational capabilities observed across different types of QRC systems present an intriguing research avenue: the investigation of quantum systems through their computational performance when utilized in the QRC. 

In this study, we propose an inverse extension of the QRC framework, termed quantum reservoir probing (QRP). 
While the QRC aims to exploit quantum systems for computational purposes, the QRP is specifically dedicated to elucidating quantum many-body physics from a computational point of view. 
Notably, in recent years, the intersection of quantum information and quantum matter has gained prominence in various contexts, highlighting the utility of quantum information in probing nonequilibrium quantum many-body phenomena such as quantum chaos~\cite{JHEP:Shenker:2014,JHEP:Maldacena:2016,JHEP:Hosur:2016,PRL:Garc:2018,PRB:Ben-Zion:2018,PRX:Chan:2018,PRX:Khemani:2018}, thermalization dynamics~\cite{Nature:Rigol:2008,NewJPhys:Bohrdt:2017,PRB:Zhou:2017,PRB:Swingle:2017,PRB:Luitz:2017,PRL:Sahu:2019,NatComm:Lewis:2019}, and dynamical quantum phase transitions~\cite{PRL:Heyl:2018,PRA:Wang:2019,PRL:Lewis:2020,PRB:Bin:2023}. 
Analogously, the QRP investigates quantum phenomena through an interdisciplinary approach by establishing a correspondence between the computational performance of the QRC and the physical attributes of the employed quantum system. 
Since a variety of phenomena can be associated with computation by judiciously selecting the information processed or the computational tasks performed, the QRP has broad applications in the exploration of quantum many-body physics. 
As a fundamental demonstration of the research avenues via the QRP, we here investigate the dynamics of information propagation within quantum systems, where locally encoded quantum information spreads over a multitude of degrees of freedom. 
Although such local information often becomes inaccessible to local probes as the dynamics progress toward the long-time limit (quantum information scrambling~\cite{JHEP:Patrick:2007,JHEP:Sekino:2008}), our study focuses on the early timescale, far from being fully scrambled, where understanding how information is distributed in the Hilbert space at each moment becomes a pertinent question.

In this application of the QRP, information is directly monitored analogously to a pump-probe paradigm. 
Random information is locally injected into the quantum system under investigation, and the system's response is recorded in a selected degree of freedom. 
Subsequently, the original input value is estimated using the observation outcomes based on a statistical approach. 
Successful estimation signifies that information has propagated to that read-out degree of freedom; otherwise, it remains unpropagated therein. 
Utilizing this estimation performance as an indicator, the QRP can comprehensively assess information propagation to an arbitrary degree of freedom at arbitrary time in a unified manner. 
We demonstrate the efficacy of the QRP by investigating a one-dimensional quantum Ising chain as a paradigmatic example. 
We show that the QRP distinctly captures the information propagation dynamics that reflects the intrinsic dynamical characteristics of the system, such as quasiparticle-mediated propagation in an integrable free fermion system and correlation-mediated propagation in a quantum chaotic system. 
Moreover, by systematically scanning the read-out degrees of freedom, the QRP reveals the mechanisms governing information propagation between different degrees of freedom, namely information propagation channels, which are typically inaccessible via conventional methodologies. 
We believe that our QRP presents an interdisciplinary paradigm to further advance the understanding of quantum many-body physics.

\section{Scheme of the QRP}
\subsection{Concept of the QRP and its relationship to the QRC}
\label{sec2.1}

\begin{figure}[t!]
  \centering
  \includegraphics[width=\hsize]{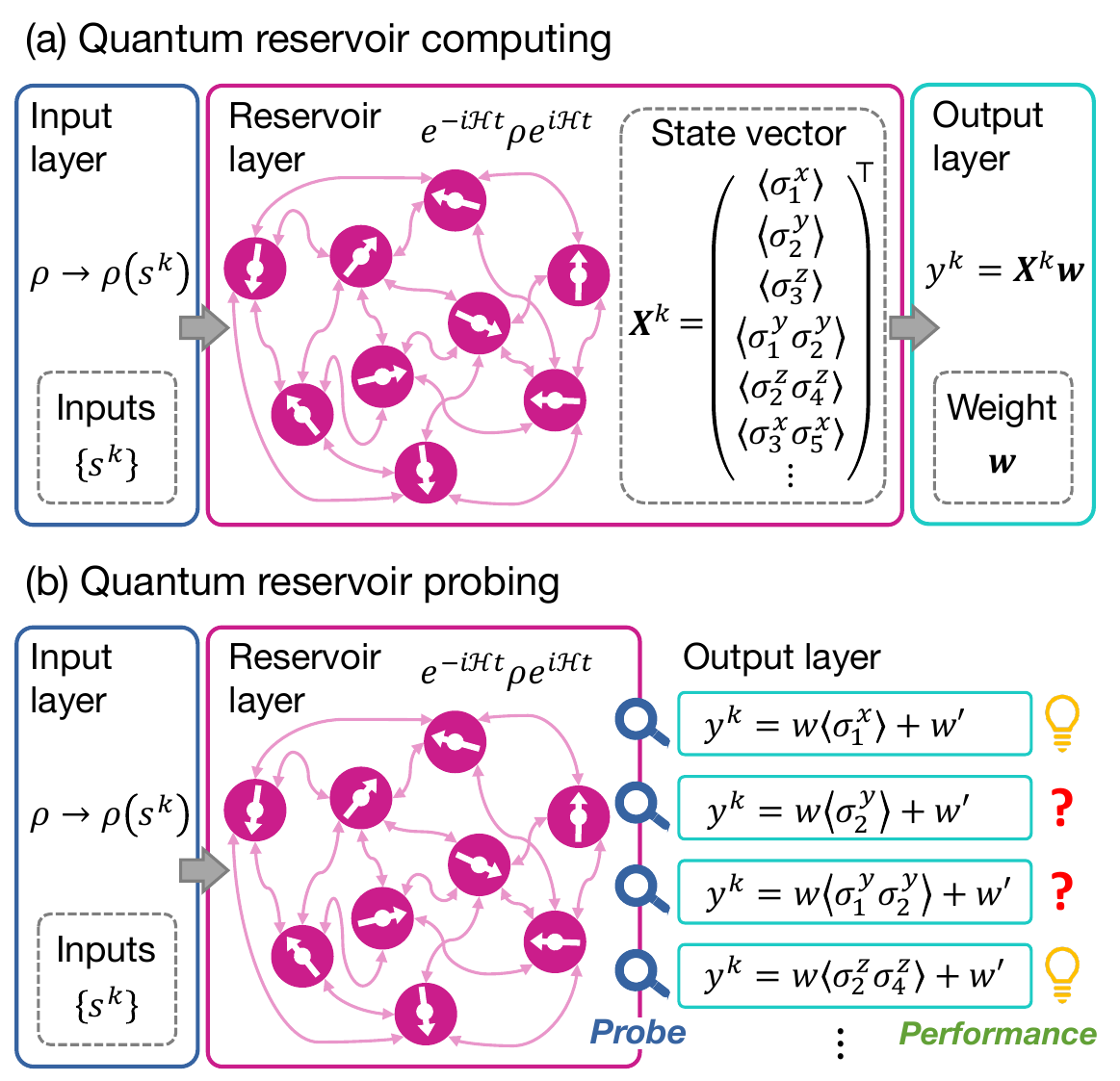}
  \caption{
    (a) Concept of the QRC. 
    Sequential inputs \(\{s^k\}\) is provided at the input layer, and the internal state of the quantum reservoir \(\bm{X}^k\) is extracted based on measurements of various degrees of freedom. 
    At the output layer, the final output \(y^k\) is computed by linearly transforming \(\bm{X}^k\) using the weight vector \(\bm{w}\). 
    (b) Schematic representation of the QRP. 
    The final output is calculated using an individual degree of freedom, whose performance elucidates the internal structure of the Hilbert space. 
  }
  \label{fig1}
\end{figure}

Prior to exploring the QRP, we introduce the QRC, a computational paradigm specifically designed to leverage quantum systems for information processing~\cite{Fujii:PRA:2017}. 
The architecture of the QRC is illustrated in Fig.~\ref{fig1}(a), comprising three layers: input, reservoir, and output. 
In the input layer, time-series input data is encoded onto a quantum system, specifically called a quantum reservoir. 
The principal role of the reservoir layer is to nonlinearly project the input data into an internal feature space, effectively emulating a network of artificial neurons with recurrent pathways. 
Unlike conventional machine learning paradigms involving optimizations, the internal attributes of the quantum reservoir remain fixed as predetermined by its inherent physical characteristics. 
This is analogous to leaving parameters within a neural network untrained, which leads to a substantial reduction in processing costs compared to schemes requiring the training of the entire weight network. 
In the reservoir layer, the dynamics of the quantum reservoir system in response to the inputs are recorded through measurements of specific variables. 
These read-out outcomes are accumulated into a one-dimensional state vector, which is then linearly transformed using a weight vector in the output layer. 
Only this weight is trained to produce the desired output for a given machine learning task. 
By leveraging the pronounced nonlinearity and high-dimensional Hilbert space of the quantum reservoir, the QRC can achieve robust neuromorphic computation solely through such simple linear post-processing. 

To harness the full potential of the quantum reservoir and access a wealth of information encoded in the Hilbert space, a straightforward approach involves measuring multiple operators, thus increasing the number of computational nodes in the post-processing stage. 
For example, when utilizing an \(N\)-site spin system as the quantum reservoir, single-site Pauli measurements \(\langle \sigma_i^\alpha \rangle\) \((1\leq j\leq N, \alpha=x,y,z)\) yield a set of \(3N\) values, and two-site Pauli measurements \(\langle \sigma_i^\alpha \sigma_j^\alpha \rangle\) \((1\leq i, j\leq N, i\neq j)\) generate a set of \(3N(N-1)/2\) values. 
Combining these measurement results, a \((3N(N+1)/2+1)\)-dimensional state vector is constructed, incorporating an additional constant term.
The inclusion of read-out outcomes from longer Pauli strings appears to enhance computational performance at first glance. 
However, empirical evidence suggests that performance tends to plateau just utilizing a number of degrees of freedom that scales polynomially with respect to \(N\)~\cite{Pena:CognComput:2023,Guillem:AdvQuantumTechnol:2023,Sannia:Quantum:2024}. 
Considering the exponentially large dimensionality of the Hilbert space, this implies that certain degrees of freedom may not contribute to computation or may extract redundant information. 
Although the selection of read-out operators is often overlooked, the suitability of a particular degree of freedom for computation should reflect the intrinsic characteristics of the Hilbert space, providing insights into the physics of the quantum reservoir system. 

The QRP is the conceptual inverse of the QRC, diverging in their primary focus: while the QRC is predominantly computationally oriented, the QRP emphasizes the underlying physical insights. 
This paradigm aligns with the recent unification of quantum information and quantum matter research, where quantum informational metrics are leveraged to unveil a variety of quantum phenomena. 
The QRP further accelerates this integration, designed to illuminate quantum many-body physics through the computational capabilities of the quantum reservoir system. 
This work demonstrates the effectiveness of the QRP in analyzing information propagation, deviating from established approaches that rely on, for example, many-body correlations, entanglement entropy, or mutual information. 

Figure \ref{fig1}(b) schematically illustrates the architecture of the QRP. 
In contrast to the QRC, our QRP enhances resolution in accessing the Hilbert space by deliberately constraining the read-out to a single degree of freedom. 
In this framework, input is supplied to and transformed within the quantum reservoir system (similar to the QRC), and the final output is calculated from the measurement outcome of a single operator (different from the QRC). 
Under this condition, the computational performance is linked to the physical attributes of the observed degree of freedom. 
Specifically, we focus on the estimation task for the input value, the performance of which quantifies the memory of input information retained in the observed degree of freedom. 
Successful (unsuccessful) estimation indicates that the input information does (does not) influence the read-out operator, thus revealing whether the information has propagated to that degree of freedom. 
For example, in Fig.~\ref{fig1}(b), the output derived from \(\langle\sigma_1^x\rangle\) exhibits superior estimation performance, suggesting the information reaches \(\sigma_1^x\) in the Hilbert space; conversely, the inferior performance obtained from \(\langle\sigma_2^y\rangle\) indicates the information does not propagate to \(\sigma_2^y\) for certain reasons. 
Analogous analysis can be applied to any degree of freedom. 
Therefore, by systematically scanning read-out operators, the QRP can evaluate information propagation in the Hilbert space with operator-level resolution. 
We note that the estimation task is meticulously selected for this research objective: by addressing alternative machine learning tasks, the QRP can probe diverse physical properties beyond information propagation.

\subsection{Formalization of the QRP to analyze information propagation}

\begin{figure}[b!]
  \centering
  \includegraphics[width=\hsize]{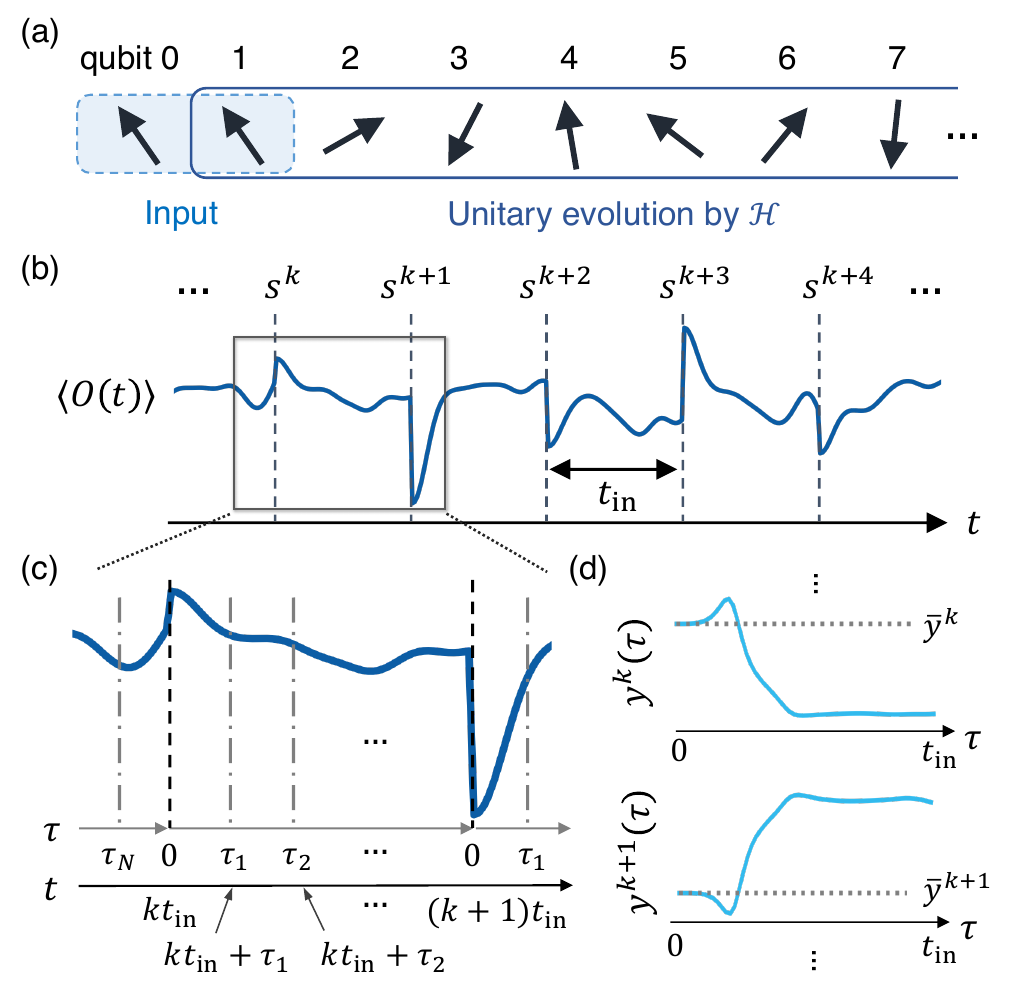}
  \caption{
    (a) Schematic of our 1D quantum spin chain. 
    Both qubit \(0\) and qubit \(1\) are simultaneously employed for input. 
    The qubits \(1, 2, \cdots\) evolve with the Hamiltonian in Eq.~(\ref{eq1}), while the qubit \(0\) is detached from the dynamics. 
    (b) Quantum dynamics of the expectation value \(\langle O(t)\rangle\) with input \(\{s^k\}\) given at time interval \(t_\mathrm{in}\). 
    (c) Concept of virtual time \(\tau\). 
    \(\langle O(\tau+kt_{\mathrm{in}})\rangle\) is used in the calculation of output \({y}^k(\tau)\). 
    (d) Dynamics of the output \({y}^k(\tau)\). The gray dotted line represents the target value \(\bar{y}^k\). The performance at \(\tau\) is evaluated based on the determination coefficient between \(\bm{y}(\tau)\) and \(\bar{\bm{y}}\).
  }
  \label{fig2}
\end{figure}

Let us formulate the QRP framework for the analysis of information propagation. 
Although we take a spin system as an illustrative example [Fig.~\ref{fig2}(a)], we emphasize that the QRP framework itself is versatile and applicable to variety of systems. 
Regarding input, we sequentially provide random input information through local quantum quenches; the QRP also accommodates alternative input methods, such as input-dependent magnetic fields or electric currents. 
Suppose \(\{s^k\}\) represent an input sequence with each \(s^k\) randomly sampled from a uniform distribution: \(s^k\in [0, 1]\). 
At every time interval \(t_{\mathrm{in}}\), the density matrix \(\rho\) is updated as
\begin{equation}
  \label{eq2}
  \rho(kt_{\mathrm{in}})\rightarrow|\psi_{\mathrm{in}}(s^k)\rangle\langle\psi_{\mathrm{in}}(s^k)|\otimes\mathrm{Tr}'\left[\rho(kt_{\mathrm{in}})\right],
\end{equation}
where \(|\psi_{\mathrm{in}}(s^k)\rangle\) represents the input state of the qubits used for encoding the information of \(s^k\), and \(\mathrm{Tr}'\) denotes the partial trace performed over the input qubits. 
Starting from the ground state, a total of \(l^{\mathrm{w}}+l^{\mathrm{tr}}+l^{\mathrm{ts}}\) inputs are provided at the input time interval  \(t_{\mathrm{in}}\). 
Among these, the initial \(l^{\mathrm{w}}\) inputs are disregarded to wash out the initial conditions, while the subsequent \(l^{\mathrm{tr}}\) and \(l^{\mathrm{ts}}\) instances are used for training and testing, respectively, as detailed below. 
Since the input procedure involves a nonunitary alteration of the quantum state, we define a virtual time \(\tau\), which is reset to zero at each input. 
Specifically, for \(kt_{\mathrm{in}}\leq t < (k+1)t_{\mathrm{in}}\), \(\tau\) is defined as \(\tau\equiv t-kt_{\mathrm{in}}\) [Fig.~\ref{fig2}(c)]. 
The system subsequently undergoes time evolution under the Hamiltonian \(\mathcal{H}\), simulated via the exact diagonalization method: \(\rho(kt_{\mathrm{in}}+\tau) = e^{-i\mathcal{H}\tau}\rho(kt_{\mathrm{in}})e^{i\mathcal{H}\tau}\). 
For read-out, the expectation value of an operator \(O\) is calculated as \(\langle O(kt_{\mathrm{in}}+\tau) \rangle = \mathrm{Tr}[\rho(kt_{\mathrm{in}}+\tau) O]\) [Fig.~\ref{fig2}(b)]. 
Hereafter, we denote \(\langle O(kt_{\mathrm{in}}+\tau)\rangle\) for general \(k\) by \(\langle O(\tau)\rangle\).

As discussed in Sec.~\ref{sec2.1}, the QRP captures information propagation through the capacity of a specific degree of freedom to estimate the input values. 
To elaborate, if information of \(s^{k}\) does not propagate to a degree of freedom \(O(\tau)\), the original value \(s^{k}\) cannot be estimated from \(\langle O(\tau)\rangle\) at all; conversely, when propagated, \(s^{k}\) can be accurately estimated from \(\langle O(\tau)\rangle\). 
Following the QRC framework~\cite{Fujii:PRA:2017}, this concept is formalized as the short-term memory (STM) task. 
The objective of this task is to produce the output \(y^k(\tau)\) that accurately estimates the target \(\bar{y}^k_d=s^{k-d}\), where \(d\) denotes the delay steps after input. 
Using the read-out \(\langle O(kt_{\mathrm{in}}+\tau)\rangle\), the estimation output at the \(k\)-th step is calculated by a linear transformation as 
\begin{equation}
  y^k_d(\tau)=w_o(\tau) \langle O(kt_{\mathrm{in}}+\tau)\rangle + w_c (\tau),\label{eq3}
\end{equation} 
where \(w_o(\tau)\) and \(w_c(\tau)\) are \(k\)-independent coefficients. 
For simplicity, we define an internal state vector 
\(\bm{X}^k(\tau) = \left(\langle O(kt_{\mathrm{in}}+\tau) \rangle, 1\right)\) and a weight vector \(\bm{w}(\tau)=(w_o(\tau),w_c(\tau))^\top\), yielding a concise representation of Eq.~(\ref{eq3}) as \(y^k_d(\tau)=\bm{X}^k(\tau)\bm{w}(\tau)\). 
The weight vector is optimized to produce the desired output using the training input dataset with \(l^{\mathrm{tr}}\) instances; subsequently, the estimation performance is evaluated on the unseen testing dataset with \(l^{\mathrm{ts}}\) instances. 

Gathering internal state vectors in the training and testing phases, we construct an \((l^{\mathrm{tr}}\times 2)\)-dimensional matrix \(X^{\mathrm{tr}}(\tau)=\{\bm{X}^k(\tau)\}_{k=l^{\mathrm{w}}+1}^{l^{\mathrm{w}}+l^{\mathrm{tr}}}\) and an \((l^{\mathrm{ts}}\times 2)\)-dimensional matrix \(X^{\mathrm{ts}}(\tau)=\{\bm{X}^k(\tau)\}_{k=l^{\mathrm{w}}+l^{\mathrm{tr}}+1}^{l^{\mathrm{w}}+l^{\mathrm{tr}}+l^{\mathrm{ts}}}\), respectively. 
The corresponding target outputs for the STM task with delay \(d\) are defined as an \(l^{\mathrm{tr}}\)-dimensional vector \(\bm{\bar{y}}^{\mathrm{tr}}_d\equiv \{s^{k-d}\}_{k=l^{\mathrm{w}}+1}^{l^{\mathrm{w}}+l^{\mathrm{tr}}} \) and an \(l^{\mathrm{ts}}\)-dimensional vector \(\bm{\bar{y}}^{\mathrm{ts}}_d\equiv \{s^{k-d}\}_{k=l^{\mathrm{w}}+l^{\mathrm{tr}}+1}^{l^{\mathrm{w}}+l^{\mathrm{tr}}+l^{\mathrm{tw}}}\). 
In the training phase, the weight vector is trained to minimize the discrepancy between the target \(\bm{\bar{y}}^{\mathrm{tr}}_d\) and the output \(\bm{y}^{\mathrm{tr}}_d(\tau)=X^{\mathrm{tr}}(\tau)\bm{w}(\tau)\) across all \(k\) at each individual moment \(\tau\) [Fig.~\ref{fig2}(d)]. 
The optimal solution minimizing the least squared error is given by 
\begin{equation}
\bm{w}(\tau)={X^{\mathrm{tr}}}^+(\tau)\bm{\bar{y}}^{\mathrm{tr}}_d,
\end{equation} 
where \({X^{\mathrm{tr}}}^+(\tau)\) denotes the Moore-Penrose pseudoinverse-matrix of \(X^{\mathrm{tr}}(\tau)\). 
In the testing phase, the estimation performance, i.e., the similarity between the target \(\bm{\bar{y}}^{\mathrm{ts}}_d\) and the testing output \(\bm{y}^{\mathrm{ts}}_d(\tau)=X^{\mathrm{ts}}(\tau)\bm{w}(\tau)\), is evaluated using the determination coefficient
\begin{equation}
  R^2_d(\tau) = \frac{\mathrm{cov}^2(\bm{y}_d^{\mathrm{ts}}(\tau),\bar{\bm{y}}^{\mathrm{ts}}_d)}{\sigma^2(\bm{y}_d^{\mathrm{ts}}(\tau))\sigma^2(\bar{\bm{y}}^{\mathrm{ts}}_d)},
\end{equation}
where \(\mathrm{cov}\) and \(\sigma^2\) represent covariance and variance, respectively. 
\(R^2_d(\tau)\) approaches one when the output \(\bm{y}_d^{\mathrm{ts}}(\tau)\) and the target \(\bar{\bm{y}}^{\mathrm{ts}}_d\) closely align; otherwise, it approaches zero. 
From this formalization of the QRP, the estimation performance functions as a quantitative metric to assess the extent to which the read-out operator \(O(\tau)\) retains the information of the \(d\)-step previous input \(s^{k-d}\). 
As shown below, our study primarily focuses on \(R^2_{d=0}(\tau)\) to elucidate the mechanisms underlying the propagation of the most recently provided information. 
Notably, since only a linear transformation is applied to the raw expectation value, the resultant performance should accurately estimate the information stored in \(O(\tau)\) without over- or under-estimation. 
Indeed, employing nonlinear transformations instead complicates the interpretation of the obtained performance, as the output reflects not only the physics associated with \(\langle O(\tau)\rangle\) but also the inherent characteristics of the chosen transformation itself.

\section{Applications: information propagation in the quantum system}
\subsection{Ballistic and diffusive dynamics of information propagation}

To demonstrate the effectiveness of the QRP, we investigate information propagation in a spin-\(1/2\) Ising chain. 
The Hamiltonian is given by
\begin{equation}
  \label{eq1}
  \mathcal{H}=-J\sum_{i=1}^{N-1}\sigma^x_i\sigma^x_{i+1}+h_x\sum_{i=1}^N\sigma^x_i+h_z\sum_{i=1}^N\sigma^z_i,
\end{equation}
with \(h_z\) and \(h_x\) representing the transverse and longitudinal magnetic fields, respectively. 
\({\sigma}_i^x\) and \({\sigma}_i^z\) are the \(x\) and \(z\) Pauli matrices at site \(i\), and \(J>0\) is the strength of the nearest-neighbor Ising interaction, which we set \(J=1\) as our energy scale. 
\(N\) denotes the number of sites in the system, excluding the qubit \(0\), which is used as a reference ancillary when considering mutual information later and therefore not involved in the time evolution [Fig.~\ref{fig2}(a)]. 
The information of \(s^k\) is provided to the quantum system by setting the state of qubits \(0\) and \(1\) as \(|\psi_{\mathrm{in}}(s^k)\rangle=\sqrt{s^k}|00\rangle_{\{01\}}+\sqrt{1-s^k}|11\rangle_{\{01\}}\), following the scheme in Eq.~(\ref{eq1}). 
We take \(N=7\), \(t_{\mathrm{in}}=5\), and \((l^{\mathrm{w}},l^{\mathrm{tr}},l^{\mathrm{ts}}) = (1000,2000,2000)\) in the following calculations. 
This model is known to be mapped to a free fermion system via the Jordan-Wigner transformation in the case of \(h_x = 0\), whereas it shows chaotic spectral statistics at  \((h_x , h_z)= (-0.5, 1.05)\)~\cite{PRL:Banuls:2011}. 
We note a finite \(h_x\) breaks the symmetry \(\sigma_i^x \leftrightarrow -\sigma_i^x\).

\begin{figure}[t!]
  \centering
  \includegraphics[width=\hsize]{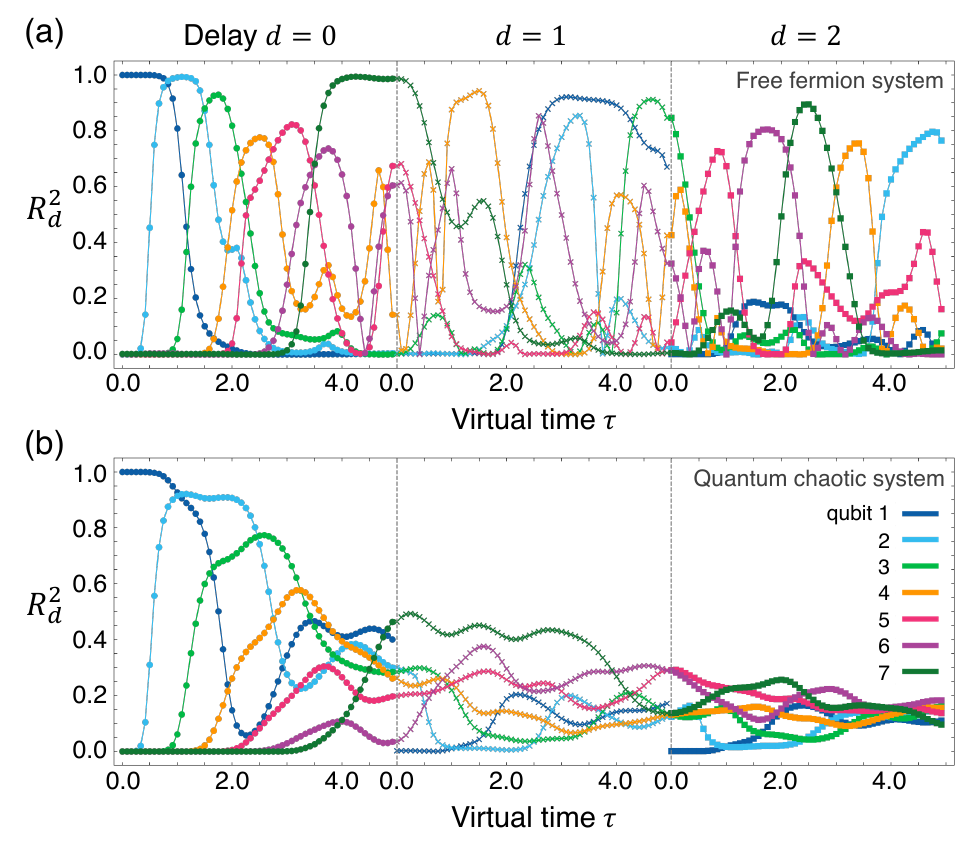}
  \caption{
    (a) Estimation performance \(R^2_d(\tau)\) in the STM task with delay \(d = 0, 1, 2\) in the free fermion system with \(h_x = 0.0\) and \(h_z = 1.0\). 
    \(\langle\sigma^z_i(\tau)\rangle\) is utilized in the calculation of the output \(y^k_d(\tau)\). 
    The colors represent each qubit, and the marker styles indicate different delays. 
    (b) The same plot as (a) in the quantum chaotic system with \(h_x = -0.5\) and \(h_z = 1.05\). 
  }
  \label{fig3}
\end{figure}

Figure \ref{fig3} represents the dynamics of the estimation performance \(R^2_d(\tau)\) for \(d=0, 1, 2\) in a free fermion system with \((h_x , h_z)= (0.0, 1.0)\) and a quantum chaotic system with \((h_x , h_z)= (-0.5, 1.05)\). 
For the calculation of the output \(y^k(\tau)\), the expectation values \(\langle\sigma^z_i(\tau)\rangle\) at each qubit \(i\) are independently employed as the read-out operator. 
In other words, \(R^2_d(\tau)\) in Fig.~\ref{fig3} quantifies the information propagated to the \(z\)-component of individual spins at each moment. 
Immediately after the input, where \(\tau \ll 1\), the information of \(s^k\) remains predominantly within the qubit \(1\) where the input is provided. 
This is evidenced by the almost unity \(R^2_{d=0}(\tau\simeq 0)\) for the qubit \(1\), while being vanishingly small for the remaining qubits. 
Subsequently, the information propagates through qubits \(2\), \(3\), \(...\), leading to a gradual emergence of nonzero values for \(R^2_{d=0}(\tau)\) from the qubits close to the qubit \(1\). 
At \(\tau = t_{\mathrm{in}}\) (\(t=(k+1)t_{\mathrm{in}}\)), the new information \(s^{k+1}\) is provided to the input qubits. 
The information of \(s^{k}\) remaining within the quantum reservoir system is then evaluated via the STM task with \(d=1\). 
Upon this input operation, the quantum state of the input qubits undergoes a substantial alteration, while the states of the other qubits remain largely unchanged. 
Indeed, \(R^2_{d-1}(\tau \rightarrow t_{\mathrm{in}})\) and \(R^2_{d}(\tau=0)\) exhibit continuous connectivity, except for the qubit 1, which is designated for input (Fig.~\ref{fig2}). 
\(R^2_{d}(\tau)\) thus effectively corresponds to \(R^2_{d=0}(\tau + t_{\mathrm{in}}d)\) under this successively quenched condition. 

Remarkably, the nature of information propagation is closely linked to dynamics of the quantum system. 
In the case of free fermion system, information propagates ballistically as illustrated in Fig.~\ref{fig3}(a). 
The dynamics of \(R^2_{d=0}(\tau)\) for each qubit exhibits a unimodal behavior, with peaks sequentially moving to neighboring sites. 
This ballistic behavior signifies quasiparticle-mediated information propagation. 
On the quenching process for inputting information, a quasiparticle containing the provided information is excited at the input qubits. 
As the quasiparticle traverses along the chain, the peak of \(R^2_{d=0}(\tau)\), representing the most recently provided information, moves to the qubit where the quasiparticle exists. 
Such localized behavior gives rise to the unimodal shape observed in Fig.~\ref{fig3}(a). 
Proceeding to the next input, a new quasiparticle is created and interferes with the existing ones. 
As a result, \(R^2_{d=1}(\tau)\) and \(R^2_{d=2}(\tau)\) exhibit relatively complicated dynamics, albeit the ballistic nature is similar to the case of \(d=0\). 

In contrast, Fig.~\ref{fig3}(b) supports diffusive information propagation in the quantum chaotic system. 
The timeline for \(R^2_{d=0}(\tau)\) commencing its ascension at each qubit is similar to that in Fig.~\ref{fig3}(a); however, the process of information accumulation toward the maximum of \(R^2_{d=0}(\tau)\) proves to be significantly prolonged. 
In addition, subsequent to reaching its maximum, \(R^2_{d=0}(\tau)\) exhibits a gradual decay over a timescale of \(t_{\mathrm{in}}\), which contrasts with an abrupt post-peak decline on a timescale of \(\Delta\tau\sim 2\) observed in the free fermion system. 
Both \(R^2_{d=1}(\tau)\) and \(R^2_{d=2}(\tau)\) similarly demonstrate smooth and gradual dynamics without any pronounced peaks. 
Notably, \(R^2_{d=2}(\tau\rightarrow t_{\mathrm{in}})\) converges toward a uniform value across all qubits, implying a homogeneous spread of information throughout all the qubits owing to the information delocalization. 
In Appendix \ref{AppA}, we present the dynamics of \(R^2_{d}(\tau)\) with varying the system size. 
Therein, the QRP captures qualitatively the same behavior as Fig.~\ref{fig3}, emphasizing the generality of the ballistic or diffusive information propagation in each system irrespective of the system size.

\begin{figure}[t!]
  \centering
  \includegraphics[width=\hsize]{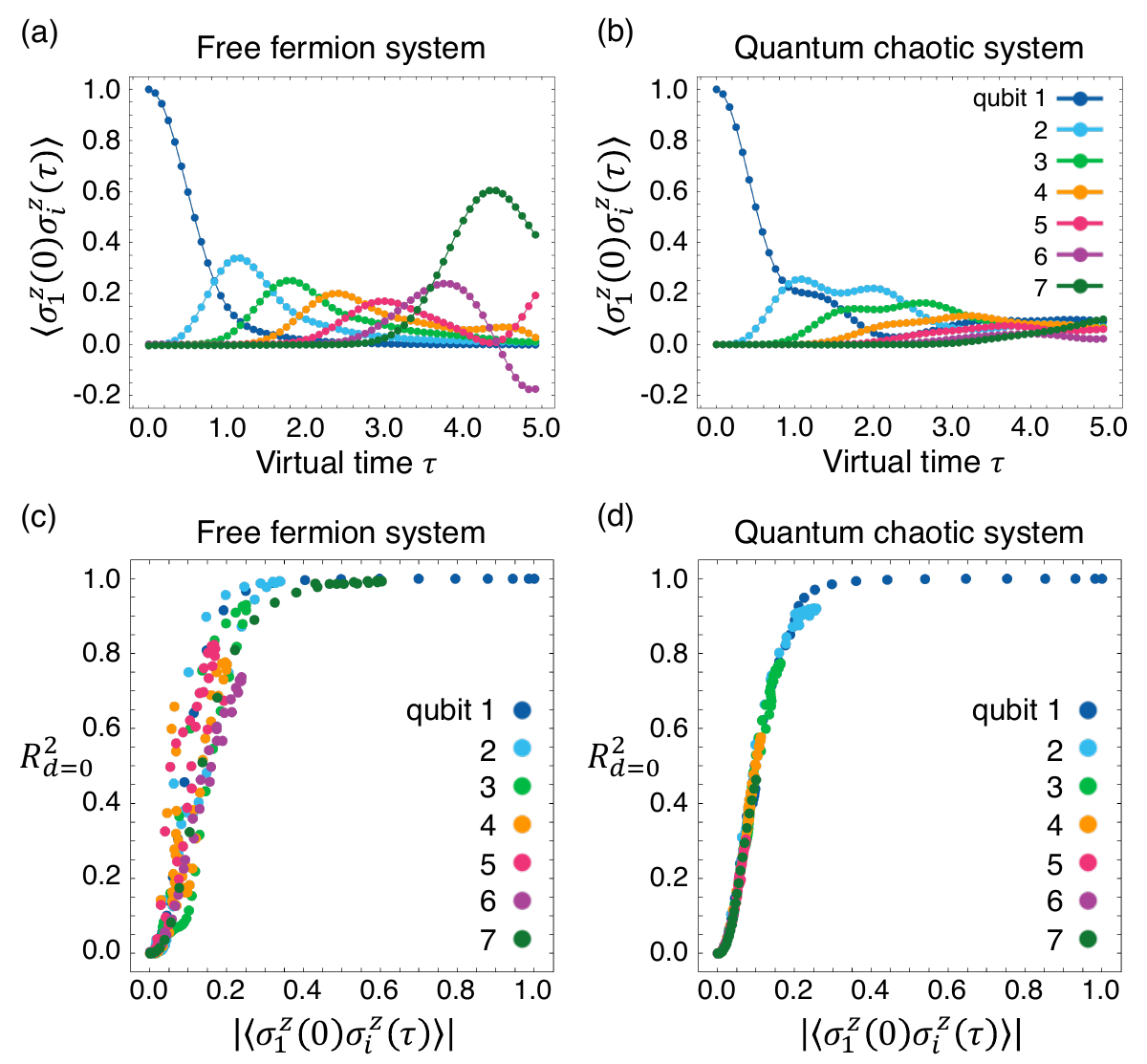}
  \caption{
    (a)-(b) The dynamical spin correlation between qubit \(1\) and each qubit \(i\) averaged over the testing inputs. 
    (c)-(d) Relationship between the dynamical spin correlation \(|\langle \sigma^z_1(0)\sigma_i^z (\tau)\rangle|\) and the estimation performance \(R^2_{d=0}(\tau)\) obtained from \(\langle\sigma_i^z\rangle\). 
    Each qubit is represented by distinct colors. 
    (a), (c) Free fermion system and (b), (d) quantum chaotic system. 
  }
  \label{fig4}
\end{figure}

For further elucidation of the mechanisms of information propagation, we examine the dynamical spin correlation between the individual qubits and the input qubit \(1\), \(\langle \sigma^z_1(0)\sigma_i^z (\tau)\rangle\), in Figs.~\ref{fig4}(a) and \ref{fig4}(b). 
In contrast to the statistically defined \(R^2_d\), physical observables, including the correlations, depend on the individual input value \(s^k\). 
Henceforth, we utilize the mean value over the testing input instances when considering physical observables. 
In the free fermion system, the dynamical spin correlation for \(i\geq 2\) initiates an ascension, achieves its maximum, and thereafter undergoes an attenuation; this entire process occurs sequentially according to the distance from qubit \(1\) [Fig.~\ref{fig4}(a)]. 
Conversely, in the quantum chaotic system in Fig.~\ref{fig4}(b), the dynamical spin correlation accumulates progressively over time and maintains a value of approximately \(0.1\) for long periods. 

Figures~\ref{fig4}(c) and \ref{fig4}(d) illustrate the relationship between the dynamical spin correlation \(|\langle \sigma^z_1(0)\sigma_i^z (\tau)\rangle|\) and the estimation performance \(R^2_{d=0}(\tau)\) when individual \(\langle\sigma_i^z (\tau)\rangle\) is employed as the read-out. 
As suggested by previous studies on classical magnetic physical reservoirs \cite{Nakane:IEEE:2018, Nakane:PRR:2021,Kobayashi:SciRep:2023}, the quantum reservoir system achieves higher \(R^2_{d=0}(\tau)\) when the spin variable harnessed as the read-out is in strong correlation with the input spin \(\sigma^z_1(0)\). 
Remarkably, despite the intricate dynamics displayed by both quantities in the quantum chaotic system [Figs.~\ref{fig3}(b) and \ref{fig4}(b)], the data collapse onto a single curve in Fig.~\ref{fig4}(d), indicating an almost one-to-one correspondence between \(|\langle \sigma^z_1(0)\sigma_i^z (\tau)\rangle|\) and \(R^2_{d=0}(\tau)\) irrespective of the qubit position \(i\) and virtual time \(\tau\). 
This is accentuated by comparison with the more dispersed plot for the free fermion system in Fig.~\ref{fig4}(c). 
We quantify the deviation from a perfect one-to-one correspondence between these two quantities using the data deviation criterion \(\Delta\). 
For a given integer \(0\leq m \leq M-1\), we define \(\Lambda_m\) as a set of \(\{(i,\tau)\}\) that satisfy \(m/M\leq |\langle \sigma_1(0)\sigma_i(\tau) \rangle | <(m+1)/M\), where \(M\) represents the number of windows (we set \(M=4,000\)). 
The average of \(R^2_{d=0}\) over \(\Lambda_m\) is denoted as \(\overline{\left(R^2_{d=0}\right)}_m\). 
Under the assumption of the one-to-one correspondence, \(R^2_{d=0}\) calculated for each \((i,\tau)\in \Lambda_m\) should exhibit be close to this average. 
The data deviation \(\Delta\) is then defined as the summation of the squared deviations from the average given by \(\Delta \equiv \sum_{m=0}^{M-1} \sum_{(i,\tau)\in\Lambda_m} \left[\left(R^2_{d=0}\right)_{(i,\tau)}-\overline{\left(R^2_{d=0}\right)}_m\right]^2\). 
In the quantum chaotic system, the deviation criterion \(\Delta\) is evaluated to be \(\Delta\simeq0.0299\), which is approximately one order of magnitude smaller compared to the value of \(\Delta \simeq 0.2866\) for the free fermion system. 
This quantitative assessment substantiates the one-to-one correspondence between the spin correlation and the estimation performance in the former system. 
In qualitative contrast to the ballistic propagation mediated by quasiparticles in the free fermion system, this observation suggests that the spin correlations play a pivotal role in the diffusive information propagation in the quantum chaotic system.

\subsection{Information propagation channels in the Hilbert space}

We here emphasize that the QRP possesses the capability to assess information propagation to any arbitrary operator \(O(\tau)\). 
The estimation performance \(R^2_d(\tau)\), derived from the output \(y(\tau)\) obtained through the linear transformation of \(\langle O(\tau)\rangle\), serves as a quantitative measure of the information stored in that degrees of freedom. 
By systematically scanning the read-out operators, the QRP can explore the spread of information across multiple degrees of freedom in the Hilbert space at any given moment, thus identifying specific channels for information propagation. 

\begin{figure}[t!]
  \centering
  \includegraphics[width=\hsize]{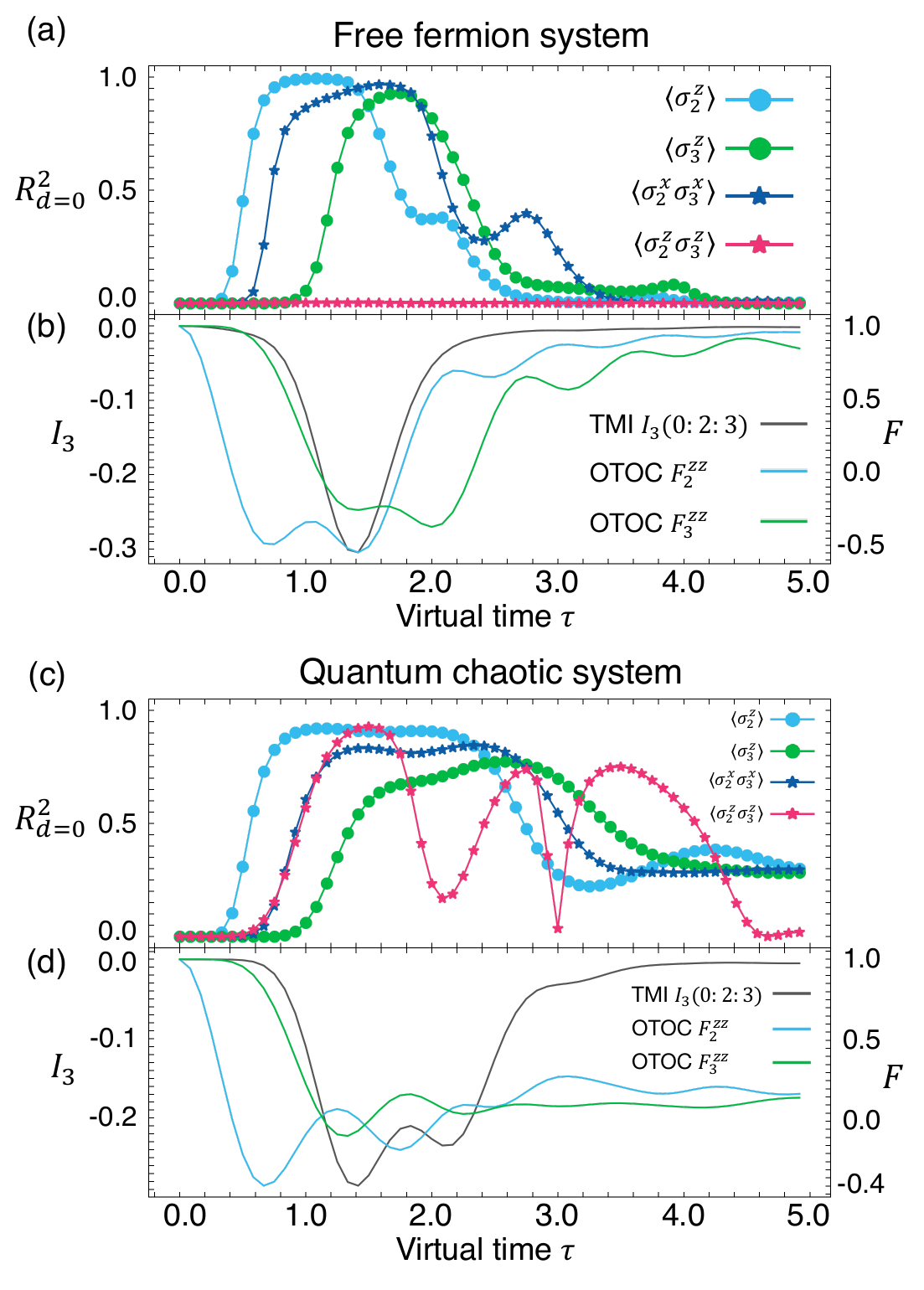}
  \caption{
    (a), (c) Dynamics of the estimation performance in the QRP framework. 
    The skyblue (green) line represents \(R^2_{d=0}(\tau)\) using single spin \(\langle \sigma_2^z (\tau)\rangle\) \(\left(\langle \sigma_3^z (\tau)\rangle\right)\) for calculation, whereas the blue (pink) line illustrates \(R^2_{d=0}(\tau)\) when the spin correlation \(\langle \sigma_2^x (\tau)\sigma_3^x (\tau)\rangle\) \(\left(\langle \sigma_2^z (\tau)\sigma_3^z (\tau)\rangle\right)\) is utilized. 
    (b), (d) Dynamics of the OTOC and TMI averaged over the testing inputs. 
    The skyblue and green lines plot the OTOC for qubit \(2\) (\(F^{zz}_2(\tau)\)) and for qubit \(3\) (\(F^{zz}_3(\tau)\)), respectively. 
    The black line displays the TMI among qubits \(0\), \(2\), and \(3\).  
    (a)-(b) Free fermion system, (c)-(d) quantum chaotic system. 
    }
  \label{fig5}
\end{figure}

Figures \ref{fig5}(a) and \ref{fig5}(c) represent \(R^2_{d=0}(\tau)\) employing observables of the single spin \(\left(\langle\sigma_2^z(\tau)\rangle\right.\) and \(\left.\langle\sigma_3^z(\tau)\rangle\right)\) and spin correlation \(\left(\langle\sigma_2^x(\tau)\sigma_3^x(\tau)\rangle\right.\) and \(\left.\langle\sigma_2^z(\tau)\sigma_3^z(\tau)\rangle\right)\); additional operators are examined in Appendix \ref{AppB}. 
In the free fermion system [Fig.~\ref{fig5}(a)], \(R^2_{d=0}(\tau)\) initially increases at the qubit \(2\), and before it becomes nonzero for the qubit \(3\), the information propagates to the \(x\) component of the correlation between the qubits \(2\) and \(3\): \(\langle\sigma_2^x(\tau)\sigma_3^x(\tau)\rangle\). 
However, \(R^2_{d=0}(\tau)\) for the \(z\) component of the correlation \(\langle\sigma_2^z(\tau)\sigma_3^z(\tau)\rangle\) remains nearly zero over all time. 
Detailed results utilizing other operators are presented in Appendix \ref{AppB}, yet it is pertinent to note that \(R^2_{d=0}(\tau)\) manifests nonzero value only when employing the \(z\) spin on a single site \(\langle\sigma^z_i(\tau)\rangle\) or the \(x\) component of the nearest-neighbor spin correlation \(\langle\sigma_i^x(\tau)\sigma_{i+1}^x(\tau)\rangle\). 
These observations unequivocally indicate that the information propagates through the channel of spin \(x\) interactions between nearest qubits. 
This is consistent with the picture of quasiparticle-mediated information propagation, as the interaction \(\sigma_i^x\sigma_{i+1}^x\) constitutes the foundation of the quasiparticle description in the free fermion system. 

In contrast, in the quantum chaotic system, both the \(x\) and \(z\) components of correlations retain information with nonzero \(R^2_{d=0}(\tau)\) [Fig.~\ref{fig5}(c)], as well as other operators, including correlations among distant qubits (Appendix \ref{AppB}). 
In particular, \(R^2_{d=0}(\tau)\) using \(\langle\sigma_2^x(\tau)\sigma_3^x(\tau)\rangle\) and  \(\langle\sigma_2^z(\tau)\sigma_3^z(\tau)\rangle\) exhibit similar behavior in the early time, and as time evolves, they diverge and display different behaviors. 
Each type of spin correlation thus serves as an individual channel for information propagation between the adjacent qubits. 
This marks a significant distinction from the free fermion system, where the information propagation channels are limited to a few correlations. 
Notably, although our investigations focus on the early time regimes, the nonzero \(R^2_d\) observed in various degrees of freedom (Appendix \ref{AppB}) could be considered as an early signature of the occurrence of quantum information scrambling in the long-time limit, where information delocalizes over diverse degrees of freedom. 

\subsection{Comparisons with OTOC and TMI}

In the previous sections, we have explored the information propagation through the estimation performance using the QRP. 
To validate its reliability, we compare the QRP with conventional methodologies for evaluating information propagation, namely the out-of-time-order correlator (OTOC) and the tripartite mutual information (TMI). 
The OTOC essentially probes the degree of noncommutativity between two initially commuting operators at different temporal points~\cite{PRL:Roberts:2015,AnnPhys:Chen:2017,JHEP:Roberts:2017,PRX:Li:2017,JHEP:Kitaev:2018,PRX:Nahum:2018,PRX:Xu:2019}. 
The long-time behavior of OTOC, particularly its asymptotic value, is a key indicator of the presence or absence of scrambling~\cite{PRB:Lin:2018,PRE:Fortes:2019,PRL:Huang:2019,PRA:Omanakuttan:2023}. 
We specifically calculate the OTOC between the qubits \(i\) and \(1\) as \(F^{zz}_i \equiv \langle \sigma^z_i(\tau)\sigma^z_1(0)\sigma^z_i(\tau)\sigma^z_1(0)\rangle\). 
On the other hand, the TMI quantifies the extent to which information about one subsystem can be extracted from the nonlocal correlations present between two other subsystems~\cite{JHEP:Hosur:2016,PRA:Iyoda:2018,PRB:Pappalardi:2018,PRA:Kuno:2022}. 
Defined in an operator-independent manner, it becomes negative when the targeted information delocalizes across the subsystems. 
Here, we utilize the detached input qubit \(0\) as the reference system for \(s^k\), and evaluate the spread of information of \(s^k\) over the nonlocal correlations between the qubits \(2\) and \(3\). 
The corresponding TMI is defined as \(I_3(0\colon2\colon3) \equiv  S_{\{0\}} +S_{\{2\}} +S_{\{3\}}- S_{\{0\}\cup\{2\}} - S_{\{0\}\cup\{3\}} - S_{\{2\}\cup\{3\}}+S_{\{0\}\cup\{2\}\cup\{3\}}\), where \(S_X\) is the von Neumann entropy. 
Both the OTOC and TMI are averaged over the testing input instances. 

Figures \ref{fig5}(b) and \ref{fig5}(d) illustrate the OTOC and the TMI in the free fermion system and the quantum chaotic system. 
In the initial stage, the OTOC \(F^{zz}_i(\tau)\) for the qubits \(2\) and \(3\) begin to decrease, slightly before the ascension of \(R^2_{d=0}(\tau)\) utilizing \(\langle\sigma^z_2(\tau)\rangle\) and \(\langle\sigma^z_3(\tau)\rangle\), respectively. 
During the intermediate stage, the TMI \(I_3(0\colon2\colon3)\) turns negative at the same time as \(R^2_{d=0}(\tau)\) calculated from the spin correlations become nonzero. 
Both of them indicate the initial spread of the input information over the qubits \(2\) and \(3\) in those time regime, which completely aligns with the behavior of \(R^2_d(\tau)\) in the QRP [Figs.~\ref{fig5}(a) and \ref{fig5}(c)]. 
As \(\tau\) approaches \(t_{\mathrm{in}}\), the OTOC \(F^{zz}_i(\tau)\) converges to \(1\) in the free fermion system and \(0\) in the quantum chaotic system. 
This long-time asymptotic value signifies the presence or absence of scrambling in each system~\cite{JHEP:Hosur:2016,PRX:Li:2017}, which is also consistent with whether or not the nonzero \(R^2_d(\tau)\) spreads for various operators in the QRP. 
These parallel observations validate the reliability of the QRP in capturing information propagation in quantum systems.

\begin{figure}[t!]
  \centering
  \includegraphics[width=\hsize]{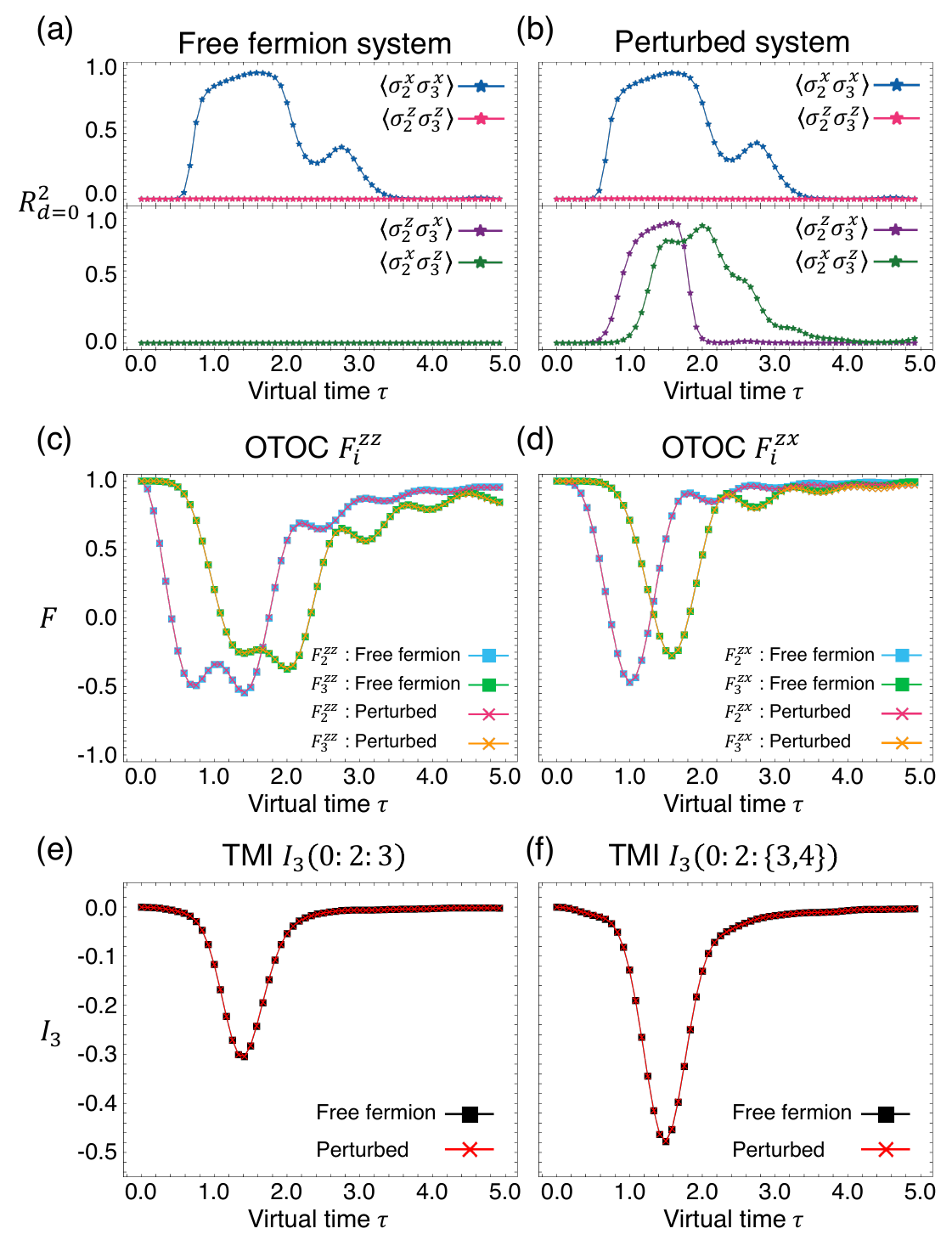}
  \caption{
    (a) The estimation performance  \(R^2_{d=0}(\tau)\) in the free fermion system with \((h_x , h_z)= (0.0, 1.0)\), employing the read-out operator \(\langle \sigma_2^x (\tau)\sigma_3^x (\tau)\rangle\) (blue), \(\langle \sigma_2^z (\tau)\sigma_3^z (\tau)\rangle\) (pink), \(\langle \sigma_2^z (\tau)\sigma_3^x (\tau)\rangle\) (purple) and \(\langle \sigma_2^x (\tau)\sigma_3^z (\tau)\rangle\) (green). 
    (b) The same plot as (a) in the perturbed system with \((h_x , h_z)= (-0.02, 1.002)\). 
    (c)-(d), The OTOC for qubit \(2\) and qubit \(3\): (c) \(F^{zz}_i = \langle \sigma^z_i(\tau)\sigma^z_1(0)\sigma^z_i(\tau)\sigma^z_1(0)\rangle\) and (d) \(F^{zx}_i = \langle \sigma^z_i(\tau)\sigma^x_1(0)\sigma^z_i(\tau)\sigma^x_1(0)\rangle\). 
    The skyblue and green lines represent the OTOC in the free fermion system, while the pink and orange lines correspond to the perturbed system. 
    (e)-(f) Dynamics of the TMI: (e) \(I_3(0\colon2\colon3)\) and (f) \(I_3(0\colon2\colon\{3,4\})\). 
    The black and red lines plot the TMI in the free fermion system and the perturbed system, respectively. 
  }
  \label{fig6}
\end{figure}

Fundamentally, the dynamics in the free fermion system and the quantum chaotic system are qualitatively disparate. 
Beyond differentiating the ballistic and diffusive propagation dynamics of \(R^2_{d=0}(\tau)\) (Fig.~\ref{fig3}), the QRP elucidates these disparities from the perspective of the information propagation channels; the pronounced distinction in \(R^2_{d=0}(\tau)\) derived from \(\langle\sigma_2^z(\tau)\sigma_3^z(\tau)\rangle\) offers compelling evidence of the differences in the propagation channels [Figs.~\ref{fig5}(a) and \ref{fig5}(c)]. 
However, such differences in propagation dynamics cannot be deduced from the behaviors of OTOC or TMI, as illustrated in Figs.~\ref{fig5}(b) and \ref{fig5}(d). 
The OTOC in these systems differ in their asymptotic values, while their early and intermediate dynamics remain notably similar, offering little insight into the information propagation channels (nor can the OTOC for other operator pairs in Appendix \ref{AppC}). 
The TMI displays qualitatively similar dynamics between these systems throughout the entire temporal regime. 
Its operator-independent definition obscures the influence of specific degrees of freedom that differentiate these quantum systems. 
Consequently, the fundamental strength of the QRP lies in its resolution to analyze the information propagation for any arbitrary degrees of freedom at any specific point in time. 
This in-depth analysis effectively uncovers the intrinsic dynamical characteristics of quantum systems, including the underlying information propagation channels. 
Moreover, it is worth highlighting the greater experimental feasibility of the QRP, as it solely requires expectation values of pertinent operators, such as spins and spin correlations. 
This stands in stark contrast to OTOC, which requires inverse time evolution, and to TMI, which necessitates highly precise quantum state tomography~\cite{PRX:Li:2017,PRL:Wei:2018,PRL:Nie:2020,NatPhys:Jochen:2022,PRL:Zhu:2022}.

We further investigate the perturbed system with \((h_x,h_z) = (-0.02,1.002)\) to lucidly demonstrate the sensitivity of the QRP. 
These parameters closely approximate those of the free fermion system; however, the system is no longer integrable, and the symmetry \(\sigma_i^x \leftrightarrow -\sigma_i^x\) is broken. 
Figures \ref{fig6}(a) and \ref{fig6}(b) show \(R^2_{d=0}(\tau)\) when each of the following is employed as the read-out operator in the free fermion system and the perturbed system, respectively: \(\langle\sigma_2^x(\tau)\sigma_3^x(\tau)\rangle\), \(\langle\sigma_2^z(\tau)\sigma_3^z(\tau)\rangle\), \(\langle\sigma_2^z(\tau)\sigma_3^x(\tau)\rangle\), and \(\langle\sigma_2^x(\tau)\sigma_3^z(\tau)\rangle\). 
Due to the similarity of the models, \(R^2_{d=0}(\tau)\) employing \(\langle\sigma_2^x(\tau)\sigma_{3}^x(\tau)\rangle\) and \(\langle\sigma_2^z(\tau)\sigma_{3}^z(\tau)\rangle\) are semiquantitatively indistinguishable between these two systems. 
Nevertheless, the breakdown of the symmetry and quasiparticle picture due to the perturbation gives rise to different types of information propagation channels beyond quasiparticle mediation, as indicated by \(R^2_{d=0}(\tau)\) utilizing \(\langle\sigma_2^z(\tau)\sigma_{3}^x(\tau)\rangle\) and \(\langle\sigma_2^x(\tau)\sigma_{3}^z(\tau)\rangle\), which only display nonzero values in the perturbed system [Fig.~\ref{fig6}(b)]. 

In Figs.~\ref{fig6}(c) and \ref{fig6}(d), we illustrate the dynamics of OTOC \(F^{zz}_i(\tau)\) and similarly defined \(F^{zx}_i(\tau)\). 
Remarkably, despite the qualitative differences between the free fermion system and the perturbed system, these OTOC manifest almost identical values in both systems, as evidenced by the overlapping pairs of curves (similar agreements are also observed for \(F^{xx}_i(\tau)\) and \(F^{xz}_i(\tau)\)). 
We also present the TMI \(I_3(0\colon2\colon3)\) and \(I_3(0\colon2\colon\{3,4\})\) in Figs.~\ref{fig6}(e) and ~\ref{fig6}(f) respectively, with the latter defined analogously to the former. 
As in the case of the OTOC, the overlapping curves therein underscore the incapacity of the TMI to distinguish between the two systems. 
These observations highlight the marked disparity in sensitivity between the QRP and the OTOC or the TMI, as the latter two exhibit less responsiveness to perturbations, even those involving changes in symmetry or integrability. 
The pronounced sensitivity of the QRP facilitates a detailed investigation of quantum many-body physics from an informational perspective, which might remain obscured in conventional analyses using the OTOC and TMI.

\section{Discussion and conclusion}

In this paper, we have proposed the QRP by inversely extending the QRC for the exploration of quantum many-body physics from the perspectives of computation and information. 
By establishing a correspondence between the physical properties and the computational performance, the QRP can shed light on the physics in any degree of freedom at arbitrary times. 
Among many applications of the QRP, we have concentrated on the study of information propagation within the Hilbert space. 
Here, sequential input information is provided to the quantum reservoir system via the local quantum quench, subsequently estimated using various read-out operators. 
The estimation performance is utilized as the metric for information propagation. 
In the quantum Ising chain with transverse and longitudinal magnetic fields, we have demonstrated that the QRP captures both ballistic information propagation mediated by quasiparticles in the free fermion system and diffusive information propagation facilitated by correlations in the quantum chaotic system, with the latter exhibiting early signatures of information scrambling across various degrees of freedom. 
Furthermore, we have shown that the QRP can systematically identify system-specific information propagation channels through a comprehensive scan of read-out operators, which is an advantage over conventional measures, in addition to its pronounced sensitivity to perturbations. 

Through the examination of information propagation, the QRP is applicable to uncover the relationship between specific operations and their resultant quantum dynamics, extending beyond the analysis of the propagation dynamics itself. 
Conventional approaches to understanding the impact of operations such as sudden quenches, application of electromagnetic fields, or coupling with external systems, typically involve direct observation of physical observables under the influence of these operations. 
However, the resulting dynamics is often affected by other multiple intrinsic and extrinsic factors, the complexity of which precludes straightforward inference of the underlying causal relationships. 
In contrast, the QRP conceptualizes quantum dynamics as a process that conveys quantum information throughout the system. 
Particularly in this study, where the input is initially provided via the quantum quench operation, the propagation of the input information can be taken as equivalent to the propagation of quenching effects, with quasiparticles or quantum correlations mediating this process. 
Similarly, when the input originates from, for example, magnetic fields or electric currents, information propagation can be interpreted as the spread of the effects of the applied fields or currents. 
Thus, by characterizing the operation as the source of the input information, the QRP can selectively extract the resultant effects in isolation from other influences of diverse origin, utilizing the provided information as a marker. 
This methodology should prove invaluable in a wide range of contexts for analyzing phenomena of interest without being obscured by the complex interplay of various factors.

Finally, we emphasize the extensive applicability of the QRP, which is not inherently limited to the analysis of information propagation. 
The QRP can investigate diverse properties of quantum systems by tailoring the input scheme and target output accordingly. 
For instance, by setting the target output as a nonlinear transformation of the provided input, the QRP would illuminate the nonlinear quantum processes within the quantum system; alternatively, by providing multiple inputs from distinct terminals, interactions among multiple excitations could be evaluated. 
Moreover, the QRP is fundamentally applicable to arbitrary systems, and potential applications to high-dimensional, dissipative, or topological systems promise to yield further insights into largely unexplored quantum many-body phenomena, including mesoscopic, non-Hermitian, and topological quantum physics. 
All these analyses could be performed within the identical framework of the QRP, which probes physics through computational performance to solve specific tasks using physical degrees of freedom. 
It is worth noting that the QRP can be implemented using the same experimental configuration as the QRC, which has already been successfully realized in several systems~\cite{Negoro:arXiv:2018,Chen:PRA:2020,Suzuki:SciRep:2022}, with potential platforms including optical lattices~\cite{Science:Choi:2016}, photonic simulators~\cite{PRL:Pierangeli:2019}, and trapped ions~\cite{NatPhys:Garttner:2017}. 
Considering the design flexibility, broad applicability, and experimental feasibility, we believe the QRP will establish itself as a potent tool for further propelling the exploration of quantum many-body physics.

\section*{Acknowledgements}
We thank Yasuyuki Kato for fruitful discussions. 
This research was supported by a Grant-in-Aid for Scientific Research on Innovative Areas ``Quantum Liquid Crystals" (KAKENHI Grant No.~JP19H05825) from JSPS of Japan and JST CREST (No. JP-MJCR18T2). 
K.K. was supported by the Program for Leading Graduate Schools (MERIT-WINGS), JSPS KAKENHI Grant Number JP24KJ0872, and JST BOOST Grant Number JPMJBS2418.

\begin{appendix}
\numberwithin{equation}{section}

\appendix

\section{Information propagation dynamics with varying system sizes}
\label{AppA}

\begin{figure*}[htbp]
  \centering
  \includegraphics[width=\hsize]{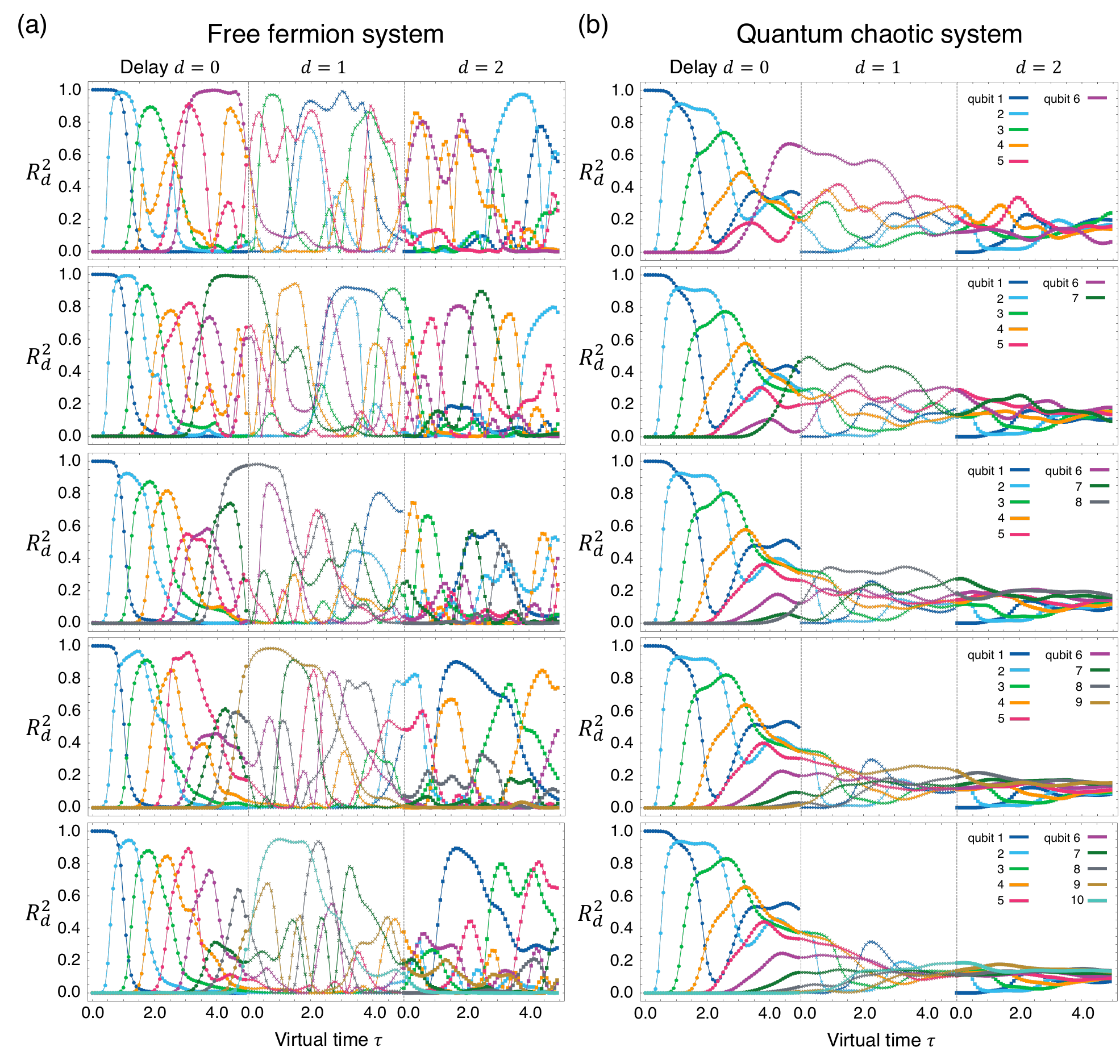}
  \caption{
    (a) Size dependence of the estimation performance \(R^2_d(\tau)\) in the STM task with delay \(d = 0, 1, 2\) in the free fermion system with \(h_x = 0.0\) and \(h_z = 1.0\). 
    \(\langle\sigma^z_i(\tau)\rangle\) is utilized in the calculation of the output \(y(\tau)\). 
    Sequentially from the top figure, the system size ranges from \(N=6\) to \(10\). 
    (b) The same plot as (a) in the quantum chaotic system with \(h_x = -0.5\) and \(h_z = 1.05\). 
    The colors represent each qubit, and the marker styles indicate different delays. 
  }
  \label{Sfig1}
\end{figure*}

We examine the system size dependence of the information propagation dynamics. 
Figure~\ref{Sfig1} extends Fig.~\ref{fig3} by showcasing the estimation performance \(R^2_d(\tau)\) calculated using individual spin operators \(\langle\sigma_i^z(\tau)\rangle\) for system sizes ranging from \(N=6\) to \(10\). 

In the free fermion system, we observe a characteristic sequential peaks in \(R^2_{d=0}\), exhibiting ballistic propagation from qubit \(1\) towards qubits \(2\), \(3\), and so forth. 
Conversely, the quantum chaotic system demonstrates diffusive propagation of \(R^2_{d=0}\) throughout the system. 
These qualitative behaviors remain consistent across different system sizes, suggesting the general applicability of the QRP for capturing the characteristics of information propagation dynamics independent of system sizes. 

\section{Read-out with various degrees of freedom}
\label{AppB}

\begin{figure*}[htbp]
  \centering
  \includegraphics[width=\hsize]{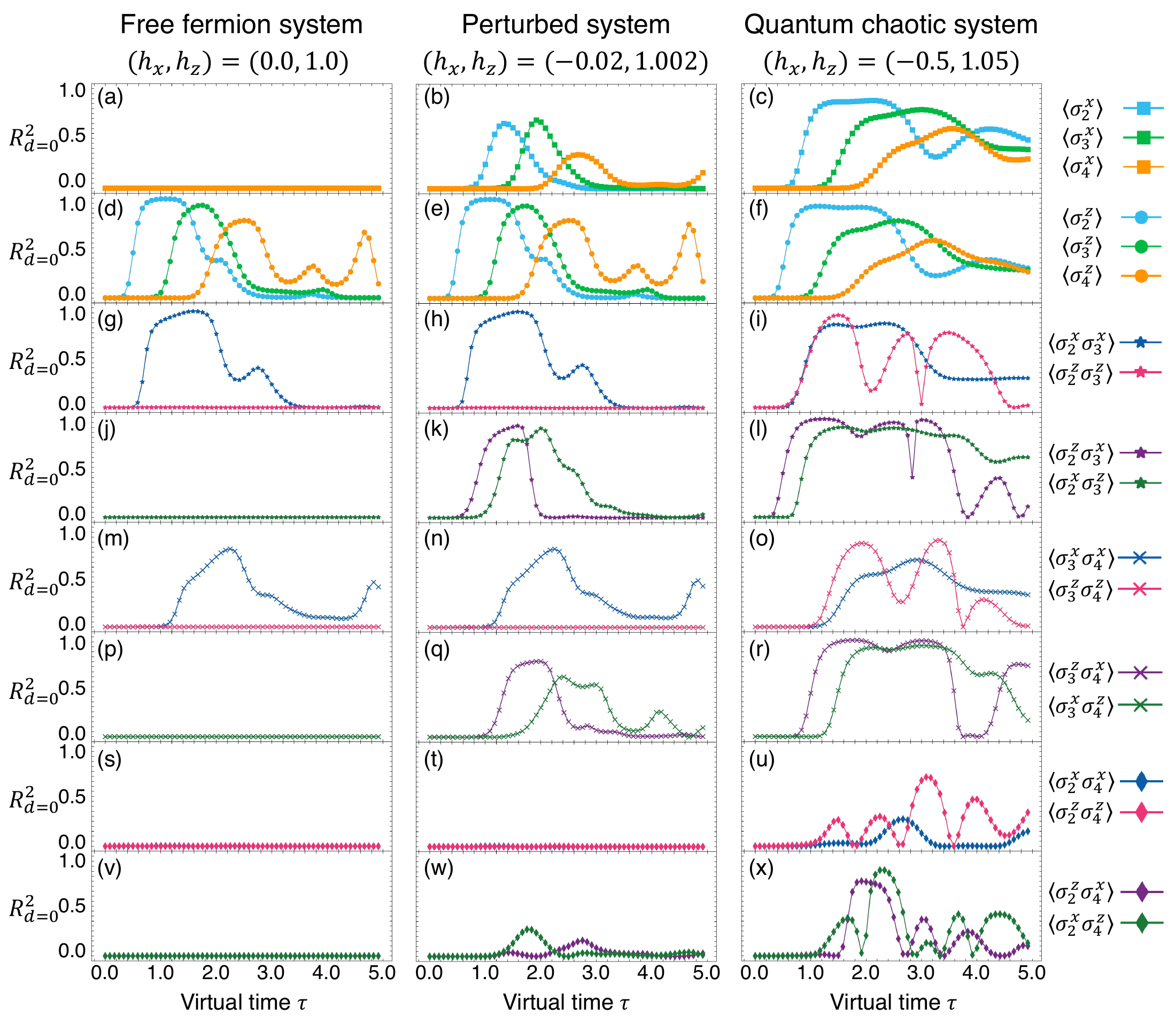}
  \caption{
    Dynamics of the reservoir performance \(R^2_{d=0}\) for the STM task employing various operators of qubits \(2,3,4\) for the read-out. 
    The legends on the rightmost side show the correspondence between markers and operators. 
    (a, d, g, j, m, p, s, v) Free fermion system, (b, e, h, k, n, q, t, w) perturbed system, and (c, f, i, l, o, r, u, x) quantum chaotic system. 
  }
  \label{Sfig2}
\end{figure*}

We investigate the information stored in various types of operators using the QRP protocol. 
Specifically focusing on the information possessed in qubits \(2\), \(3\), and \(4\), we calculate the estimation performance employing the spin \(\sigma_i^{(x,z)}(\tau)\) and spin correlations \(\sigma_i^{(x,z)}(\tau)\sigma_j^{(x,z)}(\tau)\) over these qubits. 
Figure \ref{Sfig2} illustrates \(R^2_{d=0}(\tau)\) for three different systems: the free fermion system with \((h_x,h_z) = (0.0,1.0)\), the quantum chaotic system with \((h_x,h_z) = (-0.5,1.05)\), and the perturbed system with \((h_x,h_z) = (-0.02,1.002)\). 

In the free fermion system, \(R^2_{d=0}(\tau)\) exhibits a nonzero value when either \(\langle\sigma^z_i(\tau)\rangle\) [Fig.~\ref{Sfig2}(d)] or \(\langle\sigma_i^x(\tau)\sigma_{i+1}^x(\tau)\rangle\) [Figs.~\ref{Sfig2}(g) and \ref{Sfig2}(m)] is utilized as the read-out operator. 
Otherwise, \(R^2_{d=0}(\tau)\) becomes nearly zero, indicating information is not stored in operators such as \(\langle\sigma^x_i(\tau)\rangle\) [Fig.~\ref{Sfig2}(a)], \(\langle\sigma_i^z(\tau)\sigma_{j}^z(\tau)\rangle\) [Figs.~\ref{Sfig2}(g), \ref{Sfig2}(m), and \ref{Sfig2}(s)], \(\langle\sigma_i^x(\tau)\sigma_{j}^z(\tau)\rangle\), \(\langle\sigma_i^z(\tau)\sigma_{j}^x(\tau)\rangle\) [Figs.~\ref{Sfig2}(j), \ref{Sfig2}(p), and \ref{Sfig2}(v)], and \(\langle\sigma_i^x(\tau)\sigma_{j\neq i+1}^x(\tau)\rangle\) [Fig.~\ref{Sfig2}(s)]. 
We note that due to the inherent symmetry \(\sigma_i^x \leftrightarrow -\sigma_i^x\) in the free fermion system, the expectation values of odd operators with respect to \(\sigma_i^x \) vanish, resulting in \(R^2_{d=0}(\tau)\simeq 0\) when utilizing such operators. 

In the quantum chaotic case, \(R^2_{d=0}(\tau)\) manifests nonzero values for all the read-out operators shown in Fig.~\ref{Sfig2}, including correlations between qubits \(2\) and \(4\) despite the distance between the qubits [Figs.~\ref{Sfig2}(u) and \ref{Sfig2}(x)]. 
This represents an early signature of quantum information scrambling, where information diffuses across a multitude of degrees of freedom. 

In the perturbed system, \(R^2_{d=0}(\tau)\) employing \(\langle\sigma_i^z(\tau)\rangle\) and \(\langle\sigma_i^x(\tau)\sigma_{i+1}^x(\tau)\rangle\) are semiquantitatively the same as those in the free fermion system, as evidenced by the almost identical pairs of figures: Figs.~\ref{Sfig2}(d) and \ref{Sfig2}(e), Figs.~\ref{Sfig2}(g) and \ref{Sfig2}(h), and Figs.~\ref{Sfig2}(m) and \ref{Sfig2}(n). 
However, due to the breakdown of the symmetry and the quasiparticle picture, information propagates across a broader range of degrees of freedom. 
Indeed, the \(x\) component of each spin \(\langle\sigma_i^x(\tau)\rangle\) manifests nonzero \(R^2_{d=0}(\tau)\) [Fig.~\ref{Sfig2}(b)], and the spin correlations \(\langle\sigma_i^x(\tau)\sigma_j^z(\tau)\rangle\) and \(\langle\sigma_i^z(\tau)\sigma_j^x(\tau)\rangle\) [Figs.~\ref{Sfig2}(k), \ref{Sfig2}(q), and \ref{Sfig2}(w)] also exhibit nonzero \(R^2_{d=0}(\tau)\). 
The latter suggests that these correlations serve as additional information propagation channels between qubits \(i\) and \(j\), alongside quasiparticle mediation.

\section{The OTOC for various operator pairs}
\label{AppC}

\begin{figure}[htbp]
  \centering
  \includegraphics[width=\hsize]{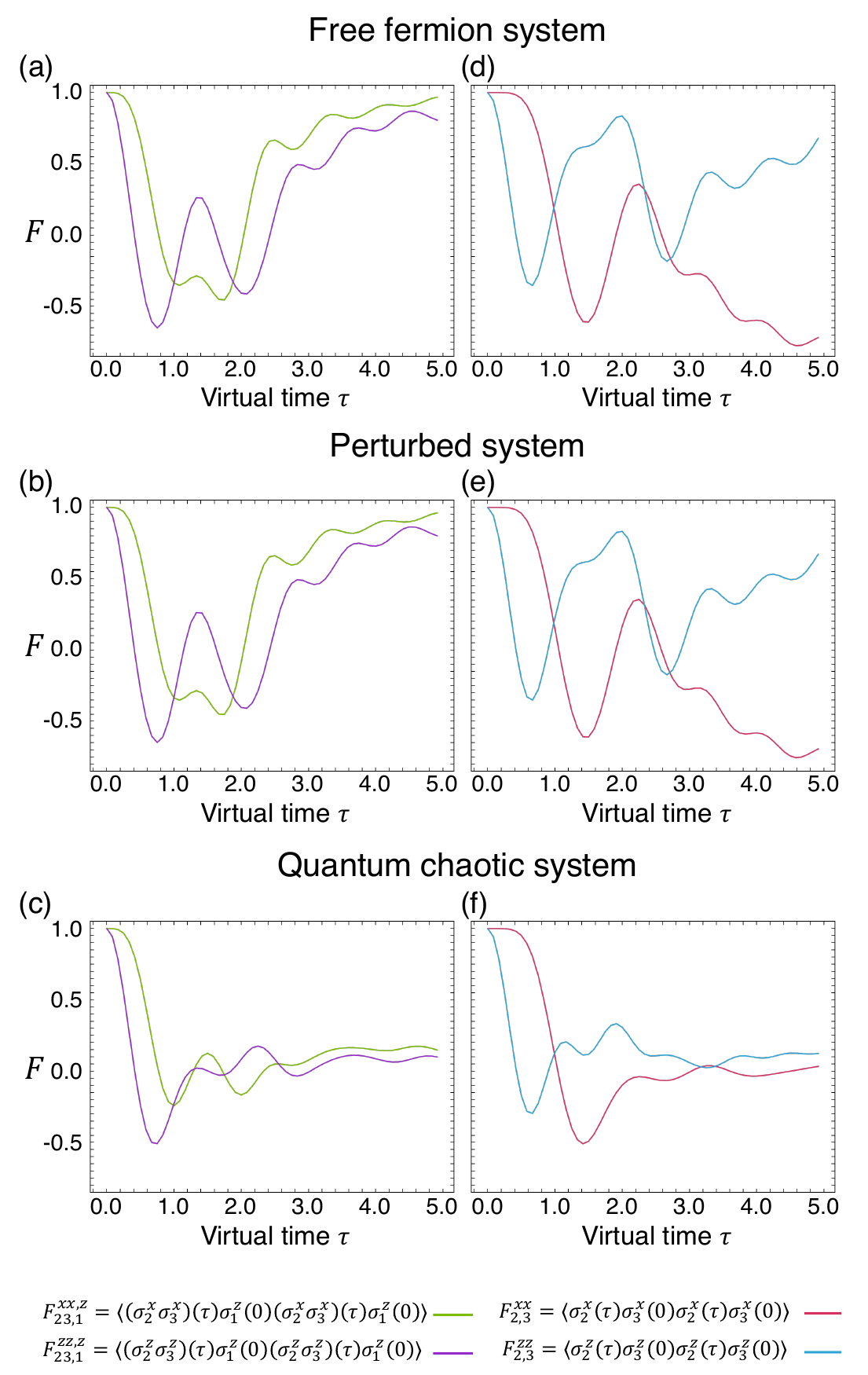}
  \caption{
    (a)-(c) Dynamics of the OTOC averaged over the testing inputs: \(F_{23,1}^{xx,z}=\langle (\sigma_2^x\sigma_3^x)(\tau)\sigma_1^z(0)(\sigma_2^x\sigma_3^x)(\tau)\sigma_1^z(0)\rangle\) (green) and \(F_{23,1}^{zz,z}=\langle (\sigma_2^z\sigma_3^z)(\tau)\sigma_1^z(0)(\sigma_2^z\sigma_3^z)(\tau)\sigma_1^z(0)\rangle\) (purple). 
    (d)-(f) The same plot for \(F_{2,3}^{xx}=\langle \sigma_2^x(\tau)\sigma_3^x(0)\sigma_2^x(\tau)\sigma_3^x(0)\rangle\) (pink) and \(F_{2,3}^{zz}=\langle \sigma_2^z(\tau)\sigma_3^z(0)\sigma_2^z(\tau)\sigma_3^z(0)\rangle\) (skyblue). 
    (a), (d) Free fermion system, (b), (e) perturbed system, and (c), (f) quantum chaotic system.
  }
  \label{Sfig3}
\end{figure}

In Fig.~\ref{fig5}(a), we demonstrate that \(R^2_{d=0}\) using \(\langle\sigma_2^x(\tau)\sigma_3^x(\tau)\rangle\) and \(\langle\sigma_2^z(\tau)\sigma_3^z(\tau)\rangle\) exhibit markedly distinct behaviors in the free fermion system, which indicates that information propagates primarily through \(\sigma_2^x(\tau)\sigma_3^x(\tau)\), rather than \(\sigma_2^z(\tau)\sigma_3^z(\tau)\). 
Conversely, in the quantum chaotic system [Fig.~\ref{fig5}(b)], \(R^2_{d=0}\) for both read-out operators becomes finite, suggesting that each type of spin correlation serves as an independent information propagation channel. 
However, the OTOC \(F_2^{zz}=\langle \sigma_2^z(\tau) \sigma_1^z(\tau) \sigma_2^z(\tau)\sigma_1^z(0) \rangle\) and \(F_3^{zz}=\langle \sigma_3^z(\tau) \sigma_1^z(\tau) \sigma_3^z(\tau)\sigma_1^z(0) \rangle\) exhibit similar dynamics between these systems, except for the asymptotic value, as shown in Figs.~\ref{fig5}(b) and \ref{fig5}(d). 
This suggests the incapability of the OTOC to identify information propagation channels in the Hilbert space.

To further illustrate this limitation, we examine the OTOC for various operator pairs, specifically focusing on the qubits \(2\) and \(3\). 
In Figs.~\ref{Sfig3}(a), \ref{Sfig3}(b), and \ref{Sfig3}(c), corresponding respectively to the free fermion system, the perturbed system, and the quantum chaotic system, we illustrate the OTOC between the qubit \(1\) and the correlations of the qubits \(2\) and \(3\): \(F_{23,1}^{xx,z}=\langle (\sigma_2^x\sigma_3^x)(\tau)\sigma_1^z(0)(\sigma_2^x\sigma_3^x)(\tau)\sigma_1^z(0)\rangle\) and \(F_{23,1}^{zz,z}=\langle (\sigma_2^z\sigma_3^z)(\tau)\sigma_1^z(0)(\sigma_2^z\sigma_3^z)(\tau)\sigma_1^z(0)\rangle\). 
As shown in Fig.~\ref{fig6}, the OTOC for the first two systems exhibit semiquantitative similarity, while those for the quantum chaotic system deviate in the asymptotic values, corresponding to the occurrence of scrambling. 
However, other qualitative differences, particularly with regard to information propagation channels, are not inferred from them. 
Even when comparing the two OTOC within the free fermion system [Fig.~\ref{Sfig3}(a)], \(F_{23,1}^{xx,z}\) and \(F_{23,1}^{zz,z}\) display similar behavior, failing to reveal any qualitative differences between \(\sigma_2^x(\tau)\sigma_3^x(\tau)\) and \(\sigma_2^z(\tau)\sigma_3^z(\tau)\), contrary to results obtained with the QRP that suggest distinct roles for these operators. 

In addition to the aforementioned OTOC involving the qubit \(1\), we further investigate the OTOC with only the qubits \(2\) and \(3\). 
For the three systems, Figs.~\ref{Sfig3}(d)-\ref{Sfig3}(f) present the OTOC \(F_{2,3}^{xx}=\langle \sigma_2^x(\tau)\sigma_3^x(0)\sigma_2^x(\tau)\sigma_3^x(0)\rangle\) and \(F_{2,3}^{zz}=\langle \sigma_2^z(\tau)\sigma_3^z(0)\sigma_2^z(\tau)\sigma_3^z(0)\rangle\); permuting the site indices \(2\) and \(3\) in the above definition yields similar results. 
As is the case in Figs.~\ref{Sfig3}(a)-\ref{Sfig3}(c), the OTOC for the free fermion and perturbed systems are semiquantitatively the same, with qualitative differences between these two systems and the quantum chaotic system emerging only in their convergence. 
Furthermore, in the free fermion system [Fig.~\ref{Sfig3}(d)], \(F_{2,3}^{xx}\) and \(F_{2,3}^{zz}\) exhibit similar behavior, except for a sign difference in the asymptotic value. 
Consequently, the OTOC \(F_{2,3}^{xx}\) and \(F_{2,3}^{zz}\), as well as \(F_{23,1}^{xx,z}\) and \(F_{23,1}^{zz,z}\), are insufficient to distinguish the different types of information propagation channels, underscoring the significant advantage of the QRP.

\end{appendix}

\end{document}